CRIPTOCURRENCIES, FIAT MONEY, BLOCKCHAINS AND DATABASES

Jorge Barrera Ortega

Abstract: Two taxonomies of money that include cryptocurrencies are analyzed. A definition of the term cryptocurrency is given and a taxonomy of them is presented, based on how its price is fixed. The characteristics of the use of current fiat money and the operation of two-level banking systems are discussed. Cryptocurrencies are compared with fiat money and the aspects in which the latter cannot be overcome are indicated. The characteristics of blockchains and databases are described. The possible cases of use of both technologies are compared, and it is noted that blockchains, in addition to cryptocurrencies and certain records, have not yet shown their usefulness, while databases constitute the foundation of most of the automated systems in operation.

1. **INTRODUCTION**

The appearance about 10 years ago of cryptocurrencies[1] has caused great expectations. Dozens, if not hundreds, of articles, many of them purely commercial and some of the academy, are published every day about their benefits and, to a lesser extent, about their defects. Many of them state that in the not too distant future, cryptocurrencies will displace the fiat[2] money currently used universally.

Along with cryptocurrencies, decentralized ledger technology (DLT) emerged, also known as blockchains. Since its inception, many experts in information and communication technology (ICT) predicted that blockchains would revolutionize the world of computer applications, displacing even databases in many cases.

The purpose of this article is to compare both the monetary aspect (cryptocurrencies vs. fiat money in its different manifestations) and the technological aspect used to support them (blockchains vs. relational databases). Many of the articles that compare cryptocurrencies and fiat money [1], in favor of the latter, focus on demonstrating that cryptocurrencies do not fulfill the classical functions of money[3]. Given that cryptocurrencies, from a historical perspective, have just been created, and that their improvement has just begun, the author thinks that this aspect, although important, is not decisive in a comparison at the present time, but rather shows issues in which cryptocurrencies still have a way to go.

As in any comparison, the point of view from which it is done largely determines the results. Some qualities that for a certain group of people can be advantageous, can be aspects that must be prevented or, at least, should be controlled for others.

---

[1] The term cryptocurrency is frequently used as a synonym for Bitcoin, the first cryptocurrency to circulate, and therefore the type of cryptocurrency it represents. As will be seen later, the characteristics of some cryptocurrencies, existing or proposed, differ greatly from those of Bitcoin.
[2] In this work the term fiat money (fiat, be made in Latin) will be used to designate the money that is issued by the central banks of the different countries and that by law is mandatory. In some works the term fiduciary money (fiduciary, trust in Latin) is still used instead of fiat, something inherited from the time when coin issuers promised that they were backed by other assets, primarily precious metals.
[3] In general, it is accepted that the money has as functions the means of exchange, unit of account and means of hoarding.

The analysis of the monetary aspect that will be developed here starts from the fact that countries need the existence of a State for its operation, that there is a legal basis that must be respected, that people have the right to a high level of privacy, as long as this privacy does not facilitate the breach of the laws. On the other hand, it must be accepted that current and future technological development will undoubtedly modify the possibilities and preferences of the human being, and of the society as a whole.

Regarding the comparison of technologies, the scope of both will be reviewed, in order to define their main areas of application, as well as their possible integration.

**2. FIAT MONEY AND CRYPTOCURRENCIES**

As stated earlier, the term cryptocurrencies are often used as a synonym for Bitcoin and the technology it uses[4].

It is well known that there are currently more than 2300 cryptocurrencies and that their number increases daily. These cryptocurrencies are of different types, from the monetary point of view, so a possible classification of them will be presented in the first place, that allows a comparison with fiat money to be established in a more systematic way.

Although cryptocurrencies have had a great presence in the popular, commercial and even academic media in recent years, the process of their use and generalization is still very incipient. Here are some figures to illustrate this fact.

The global supply of fiat money is USD 80 trillion [1], while the market value of all cryptocurrencies according to Coinmarketcap [2] was USD 248.5 billion on November 3, 2019, that is approximately 0.31% of the world's money supply.

The market value [3] of each of the 20 largest companies worldwide is greater than that of cryptocurrencies as a whole, and just the value of the largest company is 3.7 times higher.

The daily volume of currency exchange transactions (FOREX) averaged USD 6.5 trillion in April 2019 [4], while cryptocurrencies according to [2] moved a daily volume of USD 0.035[5] trillion on November 3 of 2019, 0.53% of the April average of FOREX operations.

2.1 **Taxonomies of money**

There are several recent works [5] [6] [7] in which taxonomies of money are presented and discussed, which obviously include cryptocurrencies. These works often include graphics to illustrate the proposed classification.

---

[4] In addition to blockchain, Bitcoin uses peer-to-peer networks, a protocol for solving the problem of Byzantine generals based on "proof of work," no need to personally identify those who operate the peer-to-peer network, and use of pseudonyms to identify Bitcoins holders.

[5] The figure published in Coinmarketcap [2] is USD 0.069 trillion. This figure reflects, however, according to the calculation methodology published by that website, the sum of all cryptocurrency pairs against other cryptocurrencies or currencies, while the BIS methodology only takes half of that figure. For this reason, the value presented in this work as a daily volume is half of what is published by Coinmarketcap

Below, two graphics called "the flower of money" and "the money trees" are shown, that are quite illustrative and will help us in the classification that will be presented later.

The taxonomy presented in "the flower of money" starts from four classification criteria which, at their intersections, define different types of money, namely:

• Universal accessibility

• Based on electronic procedures for its operation

• If it is issued by the Central Bank (fiat money)

• If her processing is done in a peer to peer network

Note that in this classification **cryptocurrencies** are defined as those that are issued and operated by electronic means in a peer-to-peer network. While those that are operated by the central bank, even when they also use electronic media and equal networks, are called **digital money from the central bank.**

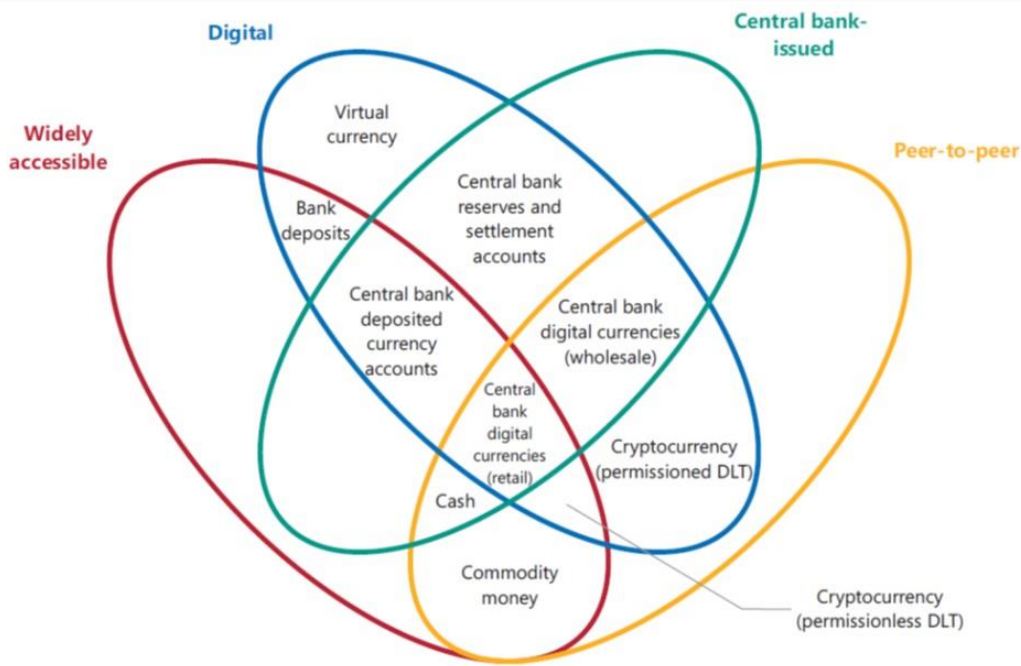

Figure 1 "the flower of money" [6]

The classification presented in Figure 2 "Money trees" is based on the following criteria:

• The type of instrument (right or object)

• The characteristics of its value (fixed, variable when reimbursed, unit of account, others)

• How it is supported in the case of rights (by the government or by the private sector)

• Technology used for its operation (decentralized or centralized)

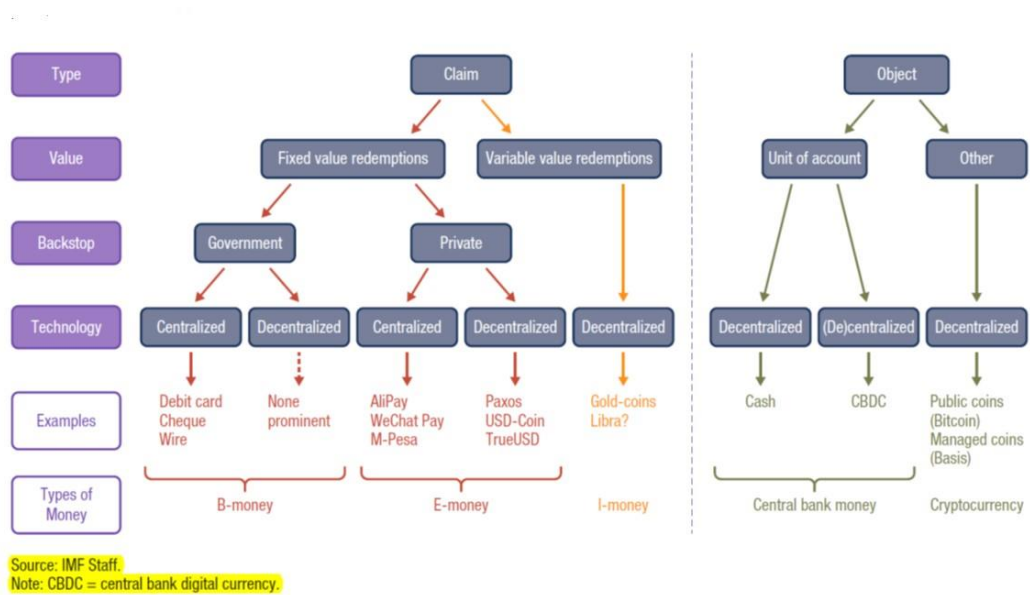

Figure 2 "Money Trees", a hierarchical taxonomy of money [5].

In "Money trees" 5 types of money are defined:

• B-money, (bank-money) i.e. debit cards, checks, transfers.

• E-money, (electronic-money) AliPay, WeChat Pay and M-Pesa are centralized examples; and Paxos, USD-Coin and TrueUSD are decentralized ones.

• I-money, (investment-money) i.e. Gold-coins and Libra.

• Central Bank money. Examples: cash and Central Bank Digital Currency.

• Cryptocurrencies. With two categories: Public currency, like Bitcoin, and managed currency, like Basis.

As can be seen in the taxonomies illustrated above, even in investigations of important institutions, different contents are used for the term cryptocurrency.

In other works [5] [8] the terms digital money and virtual money are also used to refer to some types of cryptocurrencies.

2.2 **Definition and classification of cryptocurrencies**

In this work, cryptocurrency means **any instrument that fulfills some of the functions of money; that is obtained, owned, and changed ownership by cryptographic means; and whose transactions are recorded in a blockchain.**

This definition intentionally departs from the characteristics of Bitcoin in the following ways:

• It does not necessarily have to be managed in a decentralized peer-to-peer network, in which anyone with the necessary technical resources can participate,

• It is not mandatory to protect the identity of cryptocurrency holders,

• The issuance may be centralized in the Central Bank or another financial entity,

It is understood that the novelty and universality of cryptocurrencies lies in the use of cryptography and, in particular, the construction of a blockchain for its issuance and management, something that can be done without restricting to the characteristics of Bitcoin.

Regarding the classification of cryptocurrencies that meet the definition given above, the way in which their price is fixed in relation to fiat money will be used as criteria.

This criterion is fundamental to analyze the possible use of any cryptocurrency as a means of payment of products or services, one of the functions of money that most affect the proper functioning of an economy. By law, the prices are expressed in each country using the unit of account defined by the Central Bank, which is embodied in fiat money.

The other important aspect for the use of some instrument as a means of payment is the response time necessary to accept and register a transaction. This is of utmost importance for the use of any means of payment, and in many cryptocurrencies it is not satisfactory, but in the following it will not be taken into account, since it is understood that technological development can substantially improve the current state in many cases.

Taking into account what has been explained so far, the following classification for cryptocurrencies is proposed:

• Cryptocurrencies issued by the Central Bank

Although there are several investigations in central banks for the creation of cryptocurrencies [8], cryptographic techniques and blockchains are not used even in the issuance of cash equivalents by central banks. The development of these technologies should allow in the near future their use by central banks, first, for wholesale payment systems.

• Cryptocurrencies anchored to fiat money

There are several cryptocurrencies anchored to fiat money that currently work, most with highs and lows due to lack of compliance with, or transparency in, their procedures. JPMorgan, the US financial giant, announced in February 2019 the next start-up of JPM Coin, a cryptocurrency anchored to the dollar to accelerate customer transactions. This step opens the way for the issuance by commercial banks of cryptocurrency anchored to fiat money.

• Cryptocurrencies anchored to assets (commodities or baskets of commodities or fiat currencies)

Like cryptocurrencies anchored to fiat money, there is a group of cryptocurrencies currently operating anchored to assets, with their ups and downs for similar reasons. Facebook has announced a project [9] to issue a cryptocurrency (Libra) anchored to a basket of fiat currencies, which promises to solve the problems that so far have had this type of cryptocurrency. Regardless of the financial power of Facebook, its experience in ICT and the enormous mass of costumers that use its current systems, the information available on this project does not yet allows to clearly glimpse its future.

• Cryptocurrencies whose price floats

The best known examples of this type of cryptocurrencies are Bitcoin and Ethereum. Despite its popularity in the media, due to its current difficulties, mainly related to the instability of its fiat currency prices, its use is nowadays basically speculative.

Figure 3, taken from [10], shows the monthly value of retail purchase transactions made using Bitcoin in millions of USD over a period of three and a half years until mid-2018.

As can be seen, the maximum value reached was about USD 400 million at the end of 2017, when Bitcoin got a peak of its value of the order of USD 19,000.

As stated in [10], the monthly average of the Visa card brand is USD 500 billion, 10 000 times higher than the Bitcoin average in mid-2018.

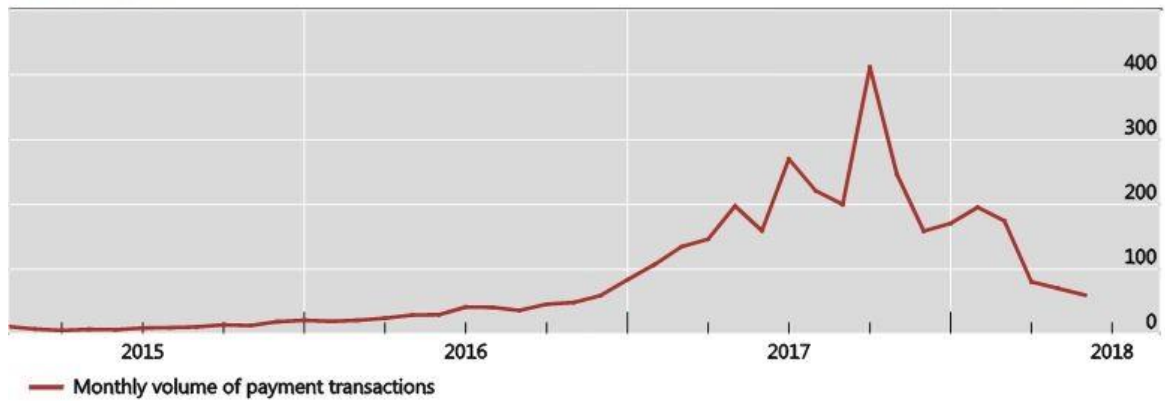

Figure 3. Monthly value of retail payments made with Bitcoins

2.3 **Special features of the issuance and use of fiat money through banking systems**

The current fiat money is the result of a development process of more than 3,000 years.

Since the creation of banking systems in a group of countries at the beginning of the 19th century, with a Central Bank that performs the functions of primary money issuance, settlement center for interbank transactions and lender of last resort, among others, the issuance of the money that exists in the whole economy is made at two levels.

• The central bank is the one who makes the primary issuance of money through cash and the acceptance of deposits from commercial banks and the government. All of these obligations of the Central Bank are called "the monetary base".

- Commercial banks[6] are authorized to take deposits and can issue money by granting loans and reflecting them in their accounting as assets before lenders and liabilities of equal magnitude that are deposited in accounts available to them. This process is called secondary money issuance, as it increases the total money available to the economy as a whole.

This secondary money issuance function is not possible in a scheme where the issuance of new funds is subject to previously defined rules and cannot be done at the discretion of third parties, as is generally the case in cryptocurrencies.

By granting loans, commercial banks not only enable the economy to be energized by providing new funds, but also play a role as promoters of efficiency, by granting them to entities that can show their ability to return them.

It is obvious that the Central Bank, directly or through bodies created for this purpose, constantly monitors the secondary money issued and sets rules to commercial banks to reduce or avoid systemic risks.

Regarding the use of fiat money, two systems for payments can be distinguished: the one for payments for high amounts, which as a rule are made by commercial banks through the, from Central Banks operated, real-time gross settlement systems (RTGSS) when different banks intervene in a transaction. When the accounts of origin and destination are in the same bank, the payments are simply annotated in the records of the bank. The other type of systems are the ones for retail payments, in which cash is used and, with increasing participation, electronic means of payment.

There is a great diversity of means of payment [11] that allows different commercial or personal situations to be resolved, in which banks play an important role in reducing the risks of these transactions, and / or helping with their financing.

An illustrative example is that of letters of credit, with which an importer can agree with an exporter to make a transaction using their respective banks. With the letter of credit, the importer's bank guarantees that the payment will be made if the exporter meets certain conditions described in that document. The exporter's bank is responsible for reviewing these conditions and, if they are met, it sends the corresponding documentation to the importer's bank, receives the payment and pays the exporter.

Another example is the possible discount of a bill of exchange accepted at term. With these operations, the financial institutions, in particular the banks, can assume a commercial credit initially granted by a seller, and advancing the amount of the bill of exchange, with the obvious discount.

---

[6] The name "bank" is given in Cuba to financial institutions that are authorized to take deposits. In other countries, there may be non-bank financial institutions that are authorized to take deposits.

## 3. BLOCKCHAINS AND DATABASES

Blockchain technology has its origin in a work signed by Satoshi Nakamoto [12] released on the internet with the name "Bitcoin: A Peer-to-Peer Electronic Cash System", and which basically describes the protocol that cryptocurrency uses today.

Current relational databases are based on a paper published in 1970 by Edgar Codd [13], where he described how to use the mathematical theory of relationships to build databases.

### 3.1 The blockchains

A blockchain is nothing more than a set of chained blocks, so that the content of a block cannot be modified without modifying the following blocks; and for this modification, certain data must be decoded for each block, with the application of cryptographic elements, which make it very difficult.

In each block transactions can be stored, or any other information whose record is to be kept without being altered.

In the original definition of blockchains, they are managed and registered in peer-to-peer computer networks, in which anyone can participate. For this mechanism to work, and for the blocks to be accepted as valid by the network, there must be a procedure that allows a consensus to be reached between the network elements, so that the so-called problem of the Byzantine generals is solved[7].

This technology is also known as "distributed ledgers technology", due to the fact that the ledgers are kept on several computers at the same time.

When blockchains are used to manage a cryptocurrency they must comply with the following principles:

• It must be ensured that the amounts of the cryptocurrency acquired by a participant are not used two or more times.

• Each transaction includes the amounts to be transferred, who transfers them and who receives them.

• Only the holder of the funds to be transferred can generate a transaction.

• The funds are stored in purses that have both a public and a private identification.

• Private identification is used to sign the transactions by the holder of the funds and the public to indicate the name of the purses of origin and destination.

• Once the transaction is registered, ownership of the funds passes to the recipient.

---

[7]The problem of the Byzantine generals is to be able to transmit a message to other generals through an enemy territory in order to coordinate an action against it, knowing that some of the generals can be traitors and that if most generals do not act together they cannot achieve victory.

• Both the transferor and the recipient of the funds can ratify in the blockchain that the transaction was recorded, so no conciliation is required.

Blockchains have been the subject of countless research in recent years that have resulted in the extension of the original concept in several ways.

3.2 **Types of blockchains**

The networks in which blockchains are managed can be of three basic types [14]:

• Permissionles: Public networks, in which anyone can participate.

• Permissioned: Consortial networks, in which several entities participate and only other entities can be incorporated if they receive a permit.

• Private: in which a single entity manages the network.

In addition to the type of network used, blockchains can be distinguished by other criteria, in particular:

• Consensus algorithm used,

• Size and characteristics of each block,

• Time needed to register a block.

There are dozens of consensus algorithms that are used in the current blockchains, each with its own particularities. In [15] the operation of the following ones is analyzed:[8]

• PoW (Proof of Work)

• PoS (Proof of Stake)

• DPoS (Delegate proof of stake)

• PBFT (Practical Byzantine Fault Tolerance)

• Ripple (Algorithm developed for the cryptocurrency Ripple)

The best-known consensus algorithm is the proof of work (PoW), that is part of the original article on Bitcoin [12].

The other algorithms are used in different cryptocurrencies. Table 1 below taken from [15] shows some properties of these algorithms.

---

[8] See also [16] that it contains a detailed description of other consensus algorithms.

| Property | PoW | PoS | DPoS | PBFT | Ripple |
| --- | --- | --- | --- | --- | --- |
| Type | Prob.-finality | Prob.-finality | Prob.-finality | Abs.-finality | Abs.-finality |
| Fault toler. | 50% | 50% | 50% | 33% | 20% |
| Power cons. | Large | Less | Less | Negligible | Negligible |
| Scalability | Good | Good | Good | Bad | Good |
| Application | Public | Public | Public | Permissioned | Permissioned |

Table 1 Some properties of selected consensus algorithms

The size and characteristics of each block don´t influence to much the efficiency of a blockchain since, in general, the process to achieve consensus is the most time-consuming, and does not directly depend on the size or characteristics of each block.

As for the time required to register each block, depending on the protocol it can be fixed, as in the case of Bitcoin, that normally takes 10 minutes, or as fast as consensus can be achieved, as in the case of Ethereum, which right now requires between 5 and 14 seconds to register a transaction.

A characteristic of blockchains introduced by the Ethereum[9] cryptocurrency is that of smart contracts. Smart contracts are programs that are incorporated into the blockchain, which monitor certain conditions and, if any of the planned ones are met, make decisions about the transfer of funds.

There are several academic articles [17] [18] that point out deficiencies in these mechanisms, some even claim that this idea should be taken with great caution [19]. From these criteria it can be affirmed that the technology of smart contracts still has a long way to go, and that its use will be restricted to very specific cases.

3.3 **The databases**

The current relational databases are the product of a development of almost 50 years, from Codd's original work [13].

The databases currently constitute the main intangible fixed asset of the automated systems in operation in the different entities [20]. Database management systems that use the relational model (RDBMS) represent the most appropriate technology for automated business management systems (ERP), including financial ones, due to their eminently transactional character and the high degree of structuring of the information they record.[10]

The end result of a database design is a set of interrelated tables. For each table, the columns it contains must be defined, including those that make up the primary and external keys, as well as the relationships with other tables and the formal rules that all rows must meet. Tables are populated with rows, the amount of which normally varies dynamically. In certain cases, the actions to be taken when a table is updated can and should be defined too.

---

[9] Actually, Bitcoin had from the beginning the possibility of writing small programs that work as smart contracts, but very rudimentary.
[10] The primary key of a table is composed of the set of columns that uniquely identify each row in the table. For this reason the combination of values of these columns cannot be repeated. External keys are combinations of columns that represent primary keys in other tables.

There are tools to facilitate the design of databases, including the entity relationship model, first proposed by Peter P. Chen in 1976 [21], which provides a methodology to define the structure of a database and standardize it [22], and has several software implementations to support that process.

There are currently multiple RDBMS[11], which guarantee the work with transactions and comply with the A.C.I.D. test[12].

The data manipulation language that has prevails is SQL (Structured Query Language), especially since 1984, when IBM announced that it would put into operation a relational database that used it [23].

An important group of improvements have been added to the initial designs of the RDBMS in the last 40 years, which have allowed them to support the solution of increasingly complex problems. Of these, the most comprehensive seems to be the inclusion of code as part of the database, in the form of user-defined functions, stored procedures, and the so-called triggers, procedures that are executed when an update of the table occurs.

Modern programming languages basically use the object-oriented paradigm, where objects are capsules that not only represent data structures, but also include procedures that apply to those structures. In these languages, the definition of new objects can start from "inheriting" the structures and procedures of already defined objects, so that code reuse is extraordinarily enhanced.

In order to bridge the differences between objects and tables in relational databases, most current programming languages are accompanied by what are known as object-relationship mappers (ORM). The ORMs allow programmers to think only of the objects defined by themselves when coding the algorithms, while at the time of execution it is the ORM that is responsible for interacting with the corresponding database tables.

As an extension of databases, data warehouses have emerged in recent years. A data warehouse is nothing more than one or several databases, which are restructured in order to facilitate the search for information relevant to decision making.

Remember that a normal database for an ERP should prioritize the ease of real-time update in transactional mode and the issuance of established periodic reports. Therefore, its structure does not necessarily facilitate obtaining interrelated information about certain entities contained in the database.

Data warehouses obviously do not have to meet the real-time update requirements in transactional mode. Their update is carried out with a periodicity never less than one day, in processes that run autonomously on the servers, usually at low load times, and which can last many hours.

---

[11] Among the most widely used RDBMS today are ORACLE, SQL-server and Postgree, the latter is an open source one.
[12] It is called the A.C.I.D. test, by the initials of the characteristics that the database must meet (Atomicity, Consistency, Isolation and Durability).

## 4. CONCLUTIONS

About the monetary aspect, As explained above, there are current functions of fiat money, which are achieved through the existence of a two-tier banking system, which cannot be covered by cryptocurrencies that are issued and managed in a decentralized manner.

From the types of cryptocurrencies defined in this work, it is clear that cryptocurrencies issued by the central bank are one more form of fiat money, as is cash and bank deposits, so comparison of this cryptocurrencies with fiat money is not necessary.

Cryptocurrencies anchored to fiat money may in the future play an important role in payment transactions and as a means of hoarding, provided that the procedures used to guarantee their convertibility to fiat money [24] are reliable, transparent and accepted in a generalized way.

Cryptocurrencies anchored to assets, to the extent that they are reliable, transparent and generally accepted as well as those anchored to fiat money, may serve as a means of hoarding, similar to the assets to which they are anchored.

Finally, cryptocurrencies whose price floats, represent a type of asset to hoard but, at least under current conditions, a not very reliable and basically speculative one.

About the technology aspect. as we have seen in the above, blockchains fulfill a very specific objective: to record transactions or other objects indelibly; while the databases fulfill a much broader objective: to allow the storage of interrelated data to simplify their recovery and, with this, to facilitate, in the first place, the automation of the management of any entity.

There are many attempts to apply[13] blockchains in fields other than those of cryptocurrencies, but as far as the author knows, no real-life applications have been achieved, that leave no doubt about their further development, as has happened with other technologies.

The databases, on the other hand, constitute the core of most of today's large automated systems. Replacing these systems using blockchains is unthinkable at the moment.

Everything seems to indicate that blockchains will continue to be used, in the immediate future, as the base technology of cryptocurrencies, in very specific applications, such as property records, and as a complement to some applications of the current databases to increase their robustness.

## 5. BLIOGRAPHIC REFERENCES

1. Burda Michael, "The Macroeconomics of Digital Currencies in Broad Brushstrokes", ISSEM Cuba seminar on Cryptocurrencies, Havana, September 2019

2. Coinmarketcap, available at https://coinmarketcap.com, accessed November 3, 2019

---

[13] Trying to apply a technology to solve a problem, because it is well-known or fashionable, is one of the most frequent mistakes in the application of IT. The correct thing is to define the tools to use from analyzing the characteristics of the problem.

3. Pricewaterhousecoopers (pwc), "The market capitalization of the world's 100 largest companies: a record figure, with 21 billion dollars," available at https://www.pwc.es/, accessed November 3, 2019

4. Bank for International Settlement (BIS), "Triennial Central Bank Survey 2019", available at www.bis.org/statistics/rpfx19.htm, 2019

5. Adrian Tobias and Mansini-Griffoli Tommaso, "The rise of digital money", Fintech notes IMF, July 2019

6. Bank for International Settlement (BIS), "Cryptocurrencies: looking beyond the hype" BIS Annual Economic Report 2018, available at www.bis.org/publ/arpdf/ar2018e5.htm, Basel, 2018

7. Bech Morten and Garratt Rodney, "Centralbank cryptocurrencies", BIS Quarterly Review, September 2017

8. "Digital currencies issued by central banks," BIS Committee on payments and market infrastructure and Market Committee, March 2018

9. Libra association members, "an introduction to Libra", whitepaper available at https://libra.org, June 2019

10. Carstens Agustin, "Money in a digital age: 10 thoughts", Speech at the Lee Kuan Yew School of Public Policy, available at www.bis.org/review/r180321.htm, Singapore, 2018

11. Barrera Jorge, "Medios de pago", Editorial UH, ISBN 978-959-7211-36-5, Havana, 2013

12. Nakamoto Satoshi, "Bitcoin: A Peer-to-Peer Electronic Cash System" available at www.bitcoin.org, 2008

13. Codd Edgar, "A Relational Model of Data for Large Shared Data Banks," Communications of the ACM, Volume 12, Number 6, June 1970

14. Paige Cabianca, "What's the difference between Public, Private and Permissioned Blockchains?" Available at https://medium.com/nakamo-to/whats-the-difference-between-a-public-and-a-private- blockchain-c08d6d1886a0, 2018

15. Zhang Shijie, Lee Jong-Hyouk, "Analysis of the main consensus protocols of blockchain", ScienceDirect, available at www.sciencedirect.com, Korea, May 2019

16. Xiao Yang, Zhang Ning, Lou Wenjing, Hou Y. Thomas, "A Survey of Distributed Consensus Protocols for BlockchainNetworks", Washington University in St. Louis, MO, Virginia Polytechnic Institute and State University, VA, October 2019

17. Atzei Nicola, Bartoletti Massimoi, Cimoli Tiziana, "A survey of attacks on Ethereum smart contracts", International Conference on Principles of Security and Trust,  Principles of Security and Trust pp 164-186, 2017

18. Fengkie Junis, Faisal Malik Widya Prasetya, Farouq Ibrahim Lubay, Anny Kartika Sari, "A Revisit on Blockchain-based Smart Contract Technology", 2018

19. O'Hara Kieron, "Smart Contracts –Dumb Idea, The Digital Citizen IEEE Computer Society, March-April 2017

20. Barrera Jorge, "Sistemas automatizados de contabilidad", Editorial UH, ISBN 978-959-7211-78-5, Havana 2016

21. Chen Peter P., "The Entity-Relationship Model - Toward a Unified View of Data". ACM Transactions on Database Systems 1 Pag. 9-36, 1976

22. Hess Kenneth, "Top 10 Enterprise Database Systems to Consider", May 20, 2010, available at http://www.serverwatch.com/trends/article.php/3883441/Top-10-Enterprise-Database-Systems-to-Consider .htm

23. Hellerstein Joseph M. and Stonebraker Michael, "Readings in Database Systems", 4th edition, The MIT Press, Cambridge, Massachusetts, 2008

24. Pernice Ingolf, Henningsen_ Sebastian, Proskalovich Roman, Florian_ Martin, Elendner Hermann, Scheuermann Björn, "Monetary Stabilization in Cryptocurrencies - Design approaches and open questions", Weizenbaum-Institute for the Networked Society, Berlin, 2019

Author's Curriculum Synthesis

Jorge Barrera Ortega. Havana 1947. Dr. in Economic Sciences. Full professor at the University of Havana. SAD Director of Juceplan (1978 - 1986). Head of the Dept. of Development of the Computing Company of Minbas (1987 - 1989). SAD Director of the National Bank of Cuba (1990 - 1997). Director SAD, Vice President and First Vice President of the Central Bank of Cuba 1997 - 2010). Member of the Desoft Technical Advisory Group. Email barrera471009@gmail.com